# Strong Terahertz Radiation Generation via Wakefield in Collisional Plasma

Divya Singh[1,2], Hitendra K. Malik[1],

[1]*PWAPA Laboratory, Department of Physics, Indian Institute of Technology Delhi, New Delhi - 110 016, India*
[2]*Department of Physics & Electronics, Rajdhani College, University of Delhi, New Delhi -110015, India*
*email- dsingh@rajdhani.du.ac.in

**Abstract:** We show that the transverse component of the wakefield can be produced with the application of an external magnetic field that generates radiation in the THz range. For this, we carry out analytical and numerical calculations to evaluate the components of the wakefield in a realistic situation of collisional plasma. We observe that the wakefield in the direction of the external magnetic field is produced, though the components of the wakefield are observed in the directions parallel and perpendicular to the direction of propagation of the laser and the magnetic field. Based on the approach of perturbation technique under quasi-static approximation, a study is made on the magnitude of emitted radiation pulses from the wakefield for the case of lasers of super-Gaussian profile. The field of THz radiation is found to be enhanced when stronger magnetic field is applied; the same is the case for the lasers of higher index.

**Keywords:** terahertz, wakefield, collisions, plasma.

## I. INTRODUCTION

Recent progress in the field of interaction of high intense lasers with gas or solid targets has been a fascinating field of research due to its tremendous applications in laser induced fusion [1], higher harmonic generation [2], laser-plasma channelling [3], laser plasma wakefield acceleration [4]. The radiations, especially the Terahertz (THz) radiation obtained based on laser plasma interaction has applications in imaging, material characterization, topography, tomography, communication etc. The researchers have reported several types of radiations including betatron type radiation, radiation from Cherenkov wakes [5–8], X-ray emission [9] and table-top hard X-ray source [10] based on inverse Compton scattering. Kacar *et al.* [11] have investigated, both theoretically and experimentally, the emission of soft X-rays from TIN plasmas for their lithographic application; they also talked about the conversion efficiency of the mechanism. Dorranian and coworkers [12, 13, 27] have reported the generation of radiation based on laser pulse and magnetized plasma interaction. Starodubstov *et al.* [14] have also done extensive studies on the plasma based radiation generation. A relationship between the signals corresponding to the Terahertz radiation and X-rays generated from laser induced plasma on gas targets has been established by Mun *et al.* [16], showing that the main cause of THz radiation is the coherent transition radiation. Nakajima [17] has reviewed the progress of laser driven electron beam and radiation sources and mentioned their various applications.

Nowadays, the THz frequency band of electromagnetic spectrum has attracted much attention of researchers and the efforts are being put on to generate the THz radiation sources where the laser plasma interaction has vital role to play [18–19]. There are a lot of studies on laser plasma interaction, which talk about different mechanisms [26, 27]. Extremely powerful THz radiations have been achieved by interaction of chirped [20], few cycle laser pulses [21] and tailored laser pulses [22] with plasmas. In this regard, Villa *et al.* [23] have been able to shape and characterize laser pulses for high gradient accelerators and Ostermayr *et al.* [24] and Devi and Malik [25] have employed super-Gaussian (sG) pulse for laser plasma acceleration with regard to their self-focusing effect. We have also made theoretical investigations for the efficient use of sG lasers for strong THz emissions in collisional plasma based on laser beat process [26, 30]. This is understood that laser pulse profile is an important tool to control plasma dynamics and the need is to devise a technique to harness the energy of plasma waves as novel radiation source in the range of THz to X-rays or even beyond that.

Plasma wakefield may serve as a source of radiation, though a plasma wave usually cannot be converted into an electromagnetic wave directly because of their different dispersion relations. However, numerical simulations reveal that intense radiation around the plasma frequency can be produced from the wakefield in inhomogeneous plasma [29]. Above mentioned all schemes talk about wakefield excitation but none of them have discussed collisional plasma. In the present work, we show that the THz radiation can be achieved based on the transverse component of the wakefield which is produced with the application of an external magnetic field. For the sake of completeness, we also include the collisions those have been neglected by other investigators.

## II. NONLINEAR CURRENT AND EXCITATION OF WAKEFIELD

We consider the propagation of a linearly polarized super-Gaussian laser (sG index-'p') of frequency $\omega$ and wave number $k$ in the z direction in plasma of uniform density $n_0$. The field of the laser is assumed as $\vec{E} = E_0 \exp[i(kz - \omega t)]\hat{y}$ where $E_0 = E_{00} \exp\left[-\left(\frac{y}{b_w}\right)^p\right]$ represents the amplitude of sG laser for p > 2; and p = 2 represents Gaussian laser. In the expression, $b_w$ is the beam width where the amplitude of the laser field becomes $1/e$ times of its peak value. The motion of plasma electrons are completely described by momentum equation. An external magnetic field of strength $B$ is assumed to be applied in the $x$- direction.

In order to derive the expression for the wakefield, we employ reductive perturbation technique (RPT). In RPT, the physical quantities such as plasma density, electron velocity, current density etc. are expanded to their higher orders in terms of smallness parameter 'a' which is the laser strength parameter ($\equiv \frac{eE_0}{mc\omega} \ll 1$). Hence, the plasma density is expanded as $n = n^{(0)} + an^{(1)} + a^{(2)}n^{(2)}$, electron velocity as $\upsilon = \upsilon^{(0)} + a\upsilon^{(1)} + a^{(2)}\upsilon^{(2)}$ and current density as $J = J^{(0)} + aJ^{(1)} + a^{(2)}J^{(2)}$.

The equation of motion of electron fluid in magnetised collisional plasma (electron neutral collision frequency - ν) is

$$m\frac{\partial \vec{\upsilon}}{\partial t} = -e[\vec{E} + \frac{1}{c}\{\vec{\upsilon} \times \vec{B}\}] - m\nu\vec{\upsilon} \tag{1}$$

The expanded equation of motion using RPT is expressed as

$$m\frac{\partial[\vec{\upsilon}^{(0)} + a\vec{\upsilon}^{(1)} + a^{(2)}\vec{\upsilon}^{(2)}]}{\partial t} = -[e(\vec{E}^{(0)} + a\vec{E}^{(1)} + a^{(2)}\vec{E}^{(2)}) + \frac{1}{c}\{(\vec{\upsilon}^{(0)} + a\vec{\upsilon}^{(1)} + a^{(2)}\vec{\upsilon}^{(2)}) \times (\vec{B}^{(0)} + a\vec{B}^{(1)} + a^{(2)}\vec{B}^{(2)})\}] - m\{(\nu^{(0)} + a\nu^{(1)} + a^{(2)}\nu^{(2)}) \times (\vec{\upsilon}^{(0)} + a\vec{\upsilon}^{(1)} + a^{(2)}\vec{\upsilon}^{(2)})\} \tag{2}$$

The above equation of motion is solved to find out the zero and first order components of velocity. These are obtained as,

$$\upsilon_x^{(0)} = 0, \upsilon_y^{(0)} = -i\frac{eE_0}{m(\omega + i\nu)}e^{i(kz-\omega t)}, \upsilon_z^{(0)} = 0, \tag{3a}$$

$$\upsilon_x^{(1)} = 0, \upsilon_y^{(1)} = \frac{eE_0(\omega + i\nu)}{im[(\omega + i\nu)^2 - \omega_c^2]}e^{i(kz-\omega t)}, \upsilon_z^{(1)} = -\frac{eE_0\omega_c}{m[(\omega + i\nu)^2 - \omega_c^2]}e^{i(kz-\omega t)}. \tag{3b}$$

Here the cyclotron frequency is represented as $\omega_c = \frac{eB}{mc}$. Now under the influence of external magnetic field in the presence of collisions, the electron oscillations become nonlinear, generating nonlinear density perturbations. Different order components of nonlinear density are computed using equation of continuity and $n^{(0)} = n_0$ where $n_0$ is the plasma ambient density. The equation of continuity is reproduced as

$$\frac{\partial n}{\partial t} + \vec{\nabla} \cdot (n_0\vec{\upsilon}) = 0 \tag{4}$$

The expanded equation of continuity using reductive perturbative technique is (RPT) is expressed as

$$\frac{\partial[N^{(0)} + aN^{(1)} + a^{(2)}N^{(2)}]}{\partial t} + \vec{\nabla} \cdot \{(N^{(0)} + aN^{(1)} + a^{(2)}N^{(2)})(\vec{\upsilon}^{(0)} + a\vec{\upsilon}^{(1)} + a^{(2)}\vec{\upsilon}^{(2)})\} = 0$$

The following first order components of nonlinear density are obtained from the above equation

$$n_x^{(1)} = 0, \quad n_y^{(1)} = \frac{k n_0 \upsilon_y^{(1)}}{(\omega - k \vec{\upsilon}_y^{(0)})}, \quad n_z^{(1)} = \frac{k n_0 \upsilon_z^{(1)}}{\omega} \tag{5}$$

We make the use of zero order components of velocity and density for finding higher order quantities. Due to the density perturbations in plasma, coupling of plasma oscillations with density fluctuations take place that drives nonlinear plasma current density. Different components of the current density are obtained from $J_j^{(1)} = -(n_0 e \upsilon_j^{(1)} + n_j^{(1)} e \upsilon_j^{(0)})$. Hence, we find

$$J_x^{(1)} = -(n_0 e \upsilon_x^{(1)} + n_x^{(1)} e \upsilon_x^{(0)}) = 0 \tag{6a}$$

$$J_y^{(1)} = -(n_0 e \upsilon_y^{(1)} + n_y^{(1)} e \upsilon_y^{(0)}) = n_0 e^2 \frac{i\omega(\omega + i\nu)}{(\omega - k\vec{\upsilon}_y^{(0)})} \frac{E_{00}}{m[(\omega + i\nu)^2 - \omega_c^2]} \exp\left[-\left(\frac{y}{b_w}\right)^p\right] e^{i(kz - \omega t)} \tag{6b}$$

$$J_z^{(1)} = -(n_0 e \upsilon_z^{(1)} + n_z^{(1)} e \upsilon_z^{(0)}) = n_0 e^2 \frac{\omega_c E_{00}}{m[(\omega + i\nu)^2 - \omega_c^2]} \exp\left[-\left(\frac{y}{b_w}\right)^p\right] e^{i(kz - \omega t)} \tag{6c}$$

We will see in the next section that how these nonlinear current densities lead to the excitation of various components of the wakefield and thereby the emission of THz radiation in collisional magnetoactive plasma.

### III. CALCULATION OF WAKEFIELDS AND THZ GENERATION

Following time dependent Maxwell's equations are used for obtaining the expression of wakefield

$$\vec{\nabla} \times \vec{E}_w = -\frac{1}{c} \frac{\partial \vec{B}_w}{\partial t} \tag{7}$$

$$\vec{\nabla} \times \vec{B}_w = \frac{4\pi}{c} \vec{J} + \frac{1}{c} \varepsilon \frac{\partial \vec{E}_w}{\partial t} \tag{8}$$

Here $\vec{E}_w$ and $\vec{B}_w$ are the electric and magnetic fields corresponding to the wakefield. We use quasistatic approximation, i.e. we consider that the laser pulse does not evolve during its propagation in the plasma. We write all the components of Maxwell's equations in terms of transformed coordinate as $\xi = z - \upsilon_g t$, where $\upsilon_g$ is the group velocity of the laser propagating along the z-axes. We obtain

$$\frac{\partial E_z}{\partial y} - \frac{\partial E_y}{\partial \xi} = \frac{\partial B_x}{\partial \xi} \tag{9a}$$

$$\frac{\partial E_x}{\partial \xi} - \frac{\partial E_z}{\partial x} = \frac{\partial B_y}{\partial \xi} \tag{9b}$$

$$\frac{\partial E_y}{\partial x} - \frac{\partial E_x}{\partial y} = \frac{\partial B_z}{\partial \xi} \tag{9c}$$

$$\frac{\partial B_z}{\partial y} - \frac{\partial B_y}{\partial \xi} = \frac{4\pi}{c} J_x^{(1)} - \frac{\partial E_x}{\partial \xi} \tag{9d}$$

$$\frac{\partial B_x}{\partial \xi} - \frac{\partial B_z}{\partial x} = \frac{4\pi}{c} J_y^{(1)} - \frac{\partial E_y}{\partial \xi} \tag{9e}$$

$$\frac{\partial B_y}{\partial x} - \frac{\partial B_x}{\partial y} = \frac{4\pi}{c} J_z^{(1)} - \frac{\partial E_z}{\partial \xi} \tag{9f}$$

The equation of motion is now evaluated for the first order velocity components under quasistatic approximation in transformed coordinate. These are obtained as

$$\frac{\partial \upsilon_x^{(1)}}{\partial \xi} = \frac{e}{mc} E_x + \frac{\nu}{c} \upsilon_x^{(1)} \tag{10a}$$

$$\frac{\partial \upsilon_y^{(1)}}{\partial \xi} = \frac{e}{mc} E_y + \frac{\omega_c}{c} \upsilon_z^{(0)} + \frac{\nu}{c} \upsilon_y^{(1)} \tag{10b}$$

$$\frac{\partial \upsilon_z^{(1)}}{\partial \xi} = \frac{e}{mc} E_z - \frac{\omega_c}{c} \upsilon_y^{(0)} + \frac{\nu}{c} \upsilon_z^{(1)} \tag{10c}$$

The use of Eqs. (9c) and (9f) in Eq. (10c) finally gives

$$\left\{ \frac{\partial^2}{\partial \xi^2} + k_p^2 \right\} \vec{E}_z = -\frac{k_p^2 m}{e} \left[ \omega_c \upsilon_y^{(0)} - \nu \upsilon_z^{(1)} \right] \tag{11}$$

Where, $k_p^2 \equiv \frac{4\pi n_0 e^2}{mc^2}$ is recognized as the wave number corresponding to the plasma wave.

Equation (11) governs the longitudinal wakefield (the component $E_z$). However, transverse component, which is responsible for the THz radiation, is also coupled to this as per Eqs. (9). At first we find the longitudinal field by solving Eq. (11) with initial condition that $E_z = 0$ at $\xi = 0$ and $L/2$, where $L$ is the length of the laser pulse. Hence, we obtain

$$E_z = \frac{PE_0}{k_p^2}\left[1 - \cos k_p\xi - \tan\frac{k_p L}{4}\sin k_p\xi\right] \tag{12}$$

Here, $P = -\dfrac{k_p^2 \omega_c(\omega^2 - \omega_c^2 + i\nu\omega)eE_0}{m(\omega + i\nu)^2\left[(\omega + i\nu)^2 - \omega_c^2\right]}e^{i(kz-\omega t)}$, which depends on the amplitude of the laser, electron-neutral collision frequency and cyclotron frequency. The transverse components of the wakefield in terms of the horizontal wake are obtained from the following relations

$$E_x = -\frac{1}{k_p^2}\frac{\partial^2 E_z}{\partial \xi \partial x}, \quad E_y = -\frac{1}{k_p^2}\frac{(\omega - k\upsilon_y^{(0)})}{\omega}\frac{\partial^2 E_z}{\partial \xi \partial y} - \frac{m\nu}{e}\upsilon_y^{(1)} \tag{13}$$

On further solving Eq. (13) for $x$ and $y$ components of generated transverse fields, we observe that there is no field in the direction of applied magnetic field, i.e. $E_x = 0$ However, transverse component of the field yields the following THz field

$$E_{0THz} = E_y = \frac{1}{k_p^2}\frac{p}{b_w}\frac{\omega_c(\omega - k\upsilon_y^{(0)})}{\omega}\frac{(\omega^2 - \omega_c^2 + i\nu\omega)E_{00}}{(i\omega - \nu)\left[(\omega + i\nu)^2 - \omega_c^2\right]}\left[\frac{y}{b_w}\right]^{p-1}e^{-(y/b_w)^p} \times$$
$$\left[\frac{\partial}{\partial\xi}(1 - \cos k_p\xi - \tan\frac{k_p L}{4}\cos k_p\xi)\right] - \frac{m\nu}{e}\upsilon_y^{(1)} \tag{14}$$

Generally the wakefield has been used for the purpose of particle acceleration and its longitudinal component serves this purpose. People have observed that the transverse component carries small magnitude. However, we show that the utility of transverse component is significant for the THz emission and its larger magnitude in the presence of wisely chosen magnetic field and profile of the laser yields stronger THz radiation. From Eq. (14) we can obtain the position of maximum amplitude of THz radiation by differentiating it w.r.t ξ. So we put $\frac{\partial E_{0THz}}{\partial \xi} = 0$ and obtain that the THz field maximizes at ξ = − L/4.

IV. Results on Wakefield

Figures 1(a) and 1(b) show the transverse profile of the longitudinal wakefield (normalized with laser field) with the dependence on the electron-neutral collision frequency ranging from $0.05\omega_p$ to $1.5\omega_p$. It is clear that the field carries higher magnitudes when the electron-neutral collisions are less in the plasma. For example, the field attains lowest magnitude when ν = $1.5\omega_p$ and largest magnitude when ν = $0.05\omega_p$. The decay in the longitudinal wakefield due to collisions is attributed to the loss of energy and momentum involved. On the other hand, this is depicted very clearly from the figures that the normalized longitudinal wakefield attains a maximum value at a particular position of y/$b_w$. This maximum amplitude is attributed to the largest amount of nonlinear current at a particular position. The magnitude of nonlinear current strongly depends on the pulse profile of the laser pulse, collision frequency and external magnetic field. Also a comparison of figs. 1(a) and 1(b) yields that the effect of magnetic field is to enhance the amplitude of the wakefield. Hence, this seems plausible that the decay due to collisions can be supplemented with the application of appropriate strength of the external magnetic field.

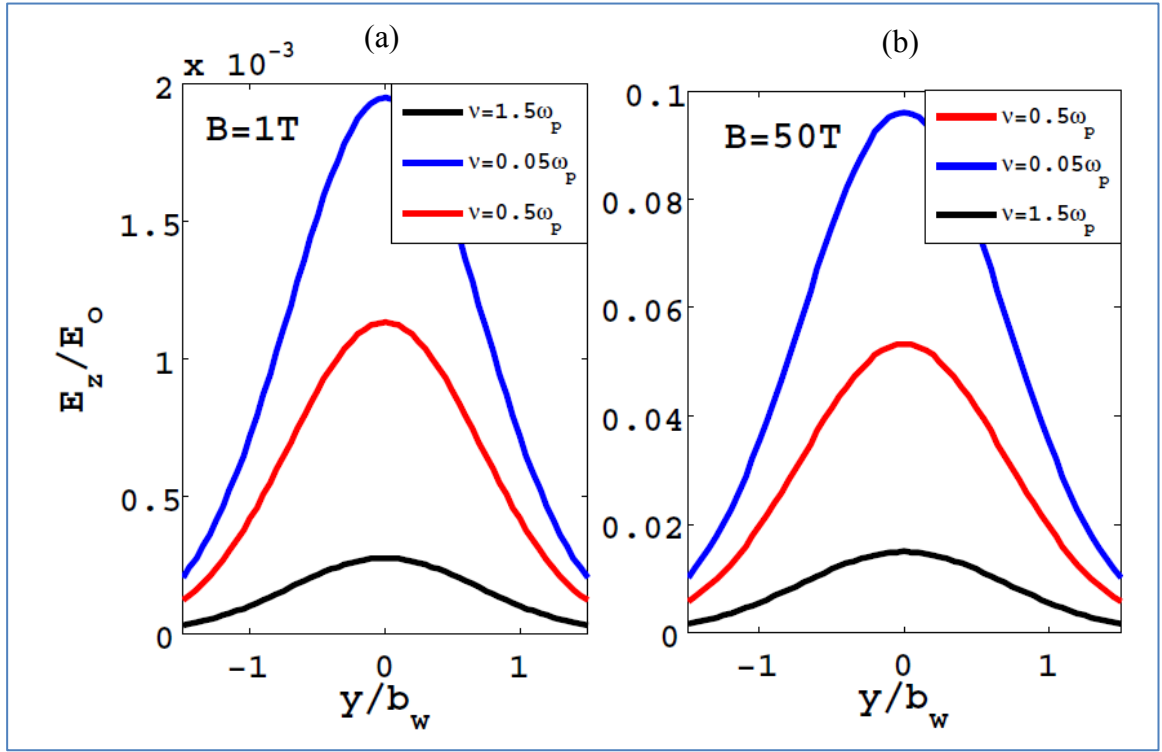

Fig.1. Transverse profile of longitudinal wakefield for various collision frequencies when p = 8, $\omega = 2.4 \times 10^{14}$ rad/sec, $\omega_p = 2.0 \times 10^{13}$ rad/sec, $L = 0.5\lambda_p$ and $\xi = -0.25L$.

Figures 2(a) and 2(b) compare the nature of wakefield obtained numerically (fig. 2a) and analytically (Fig. 2b) from Eq. (11) for different values of sG index p = 2 (black), 4 (blue) and 6 (red). It is found that there is qualitatively a close resemblance in the behaviour of field profiles obtained numerically as well as analytically, but the field amplitudes differ a bit. This figure enables us to see the results for the Gaussian (p = 2) and super-Gaussian (p > 2) lasers. This is interesting that the field amplitudes for Gaussian laser (p = 2) are much lower than the amplitudes obtained with the help of sG laser. Therefore, sG lasers are more suitable for causing excitation of large amplitude wakefield due to generation of larger currents in the plasma.

Details about the profile and features of super-Gaussian lasers may be followed from our earlier work [29].

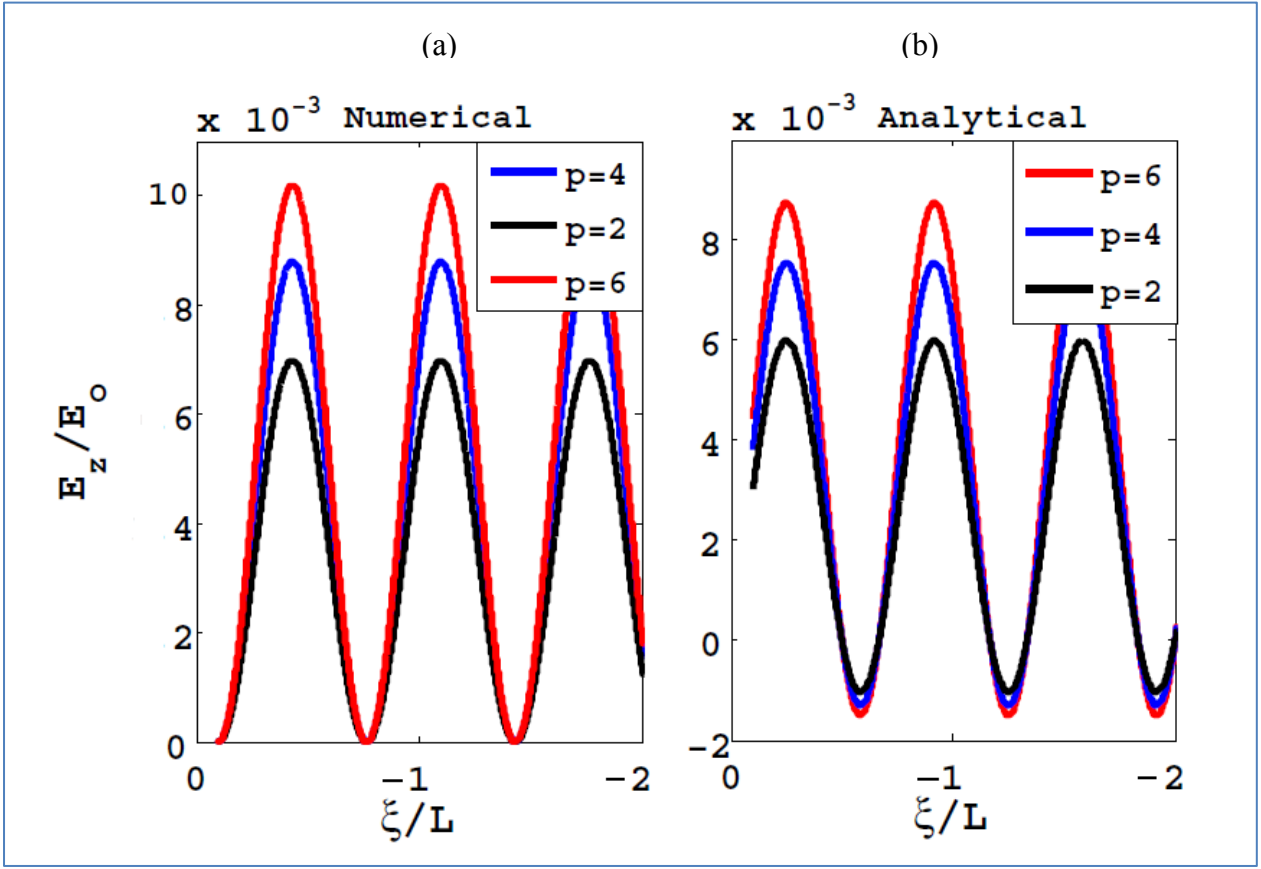

Fig.2. Comparison of wakefields obtained numerically (a) and analytically (b) when ω = 2.4×10$^{14}$ rad/sec, ω$_p$ = 2.0×10$^{13}$ rad/sec, ν = 0.05ω$_p$, B = 1T, y = 0.8b$_w$ and L = 1.5λ$_p$.

## V.  Results on THz Emission

This is seen from Eq. (14) that the amplitude of transverse field directly depends on the sG index-p and the beam width $b_w$ of the laser. Therefore, by controlling these laser parameters we can obtain higher amplitude of THz radiation. Figures 3(a) and 3(b) show the variation of normalized amplitude of emitted THz field with laser beam width for different electron-neutral collision frequency and sG index. It is clear from the figures that the THz field amplitude is decreased as the beam width is increased. This behaviour can be understood based on the sharper gradient in the laser field for small beam widths because of which larger current is expected to be generated in the plasma. On the other hand, when the collision frequency is increased from ν = 0.05$\omega_p$ to ν = 0.5$\omega_p$, the amplitude of THz field reduces significanly.

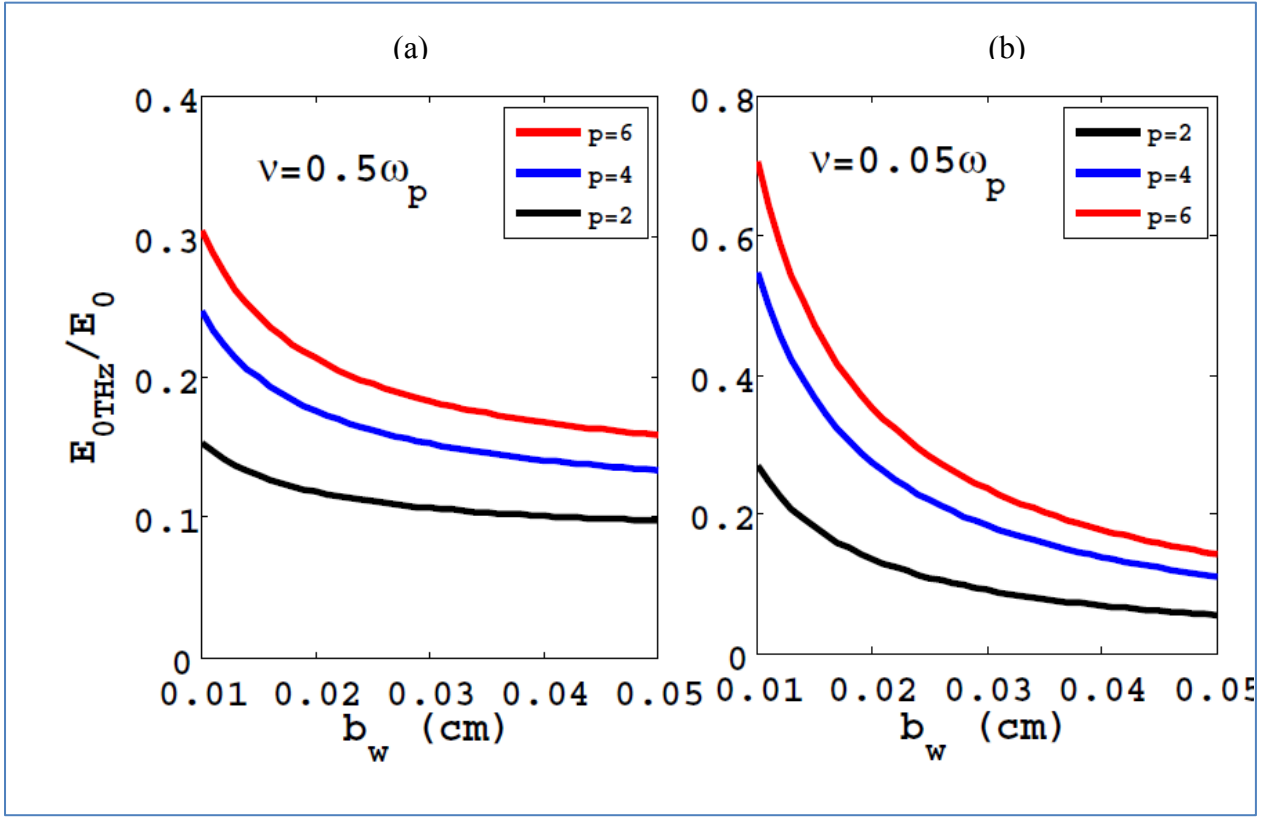

Fig.3. Variation of normalised THz field with beam width for laser of various sG index in a plasma of different collision frequency when ω = 1.15$\omega_p$, $\omega_p$ = 2.0×10$^{13}$ rad/sec , B = 1Tesla, y = 0.8$b_w$ , L = 0.5$\lambda_p$ and ξ = − 0.8L.

The impact of collision frequency on the THz field is clearly shown in fig. 4. Here, it is observed that the negative impact of ν is not so significant till ν ≈ 0.3 $\omega_p$. However, there is a drastic reduction in the field for ν > 0.3 $\omega_p$. From these results it appears that high density plasma is more suitable for the effective THz emission. Moreover, these results are consistent with the observations made by Hu *et al.* [29] and Singh and Malik [30].

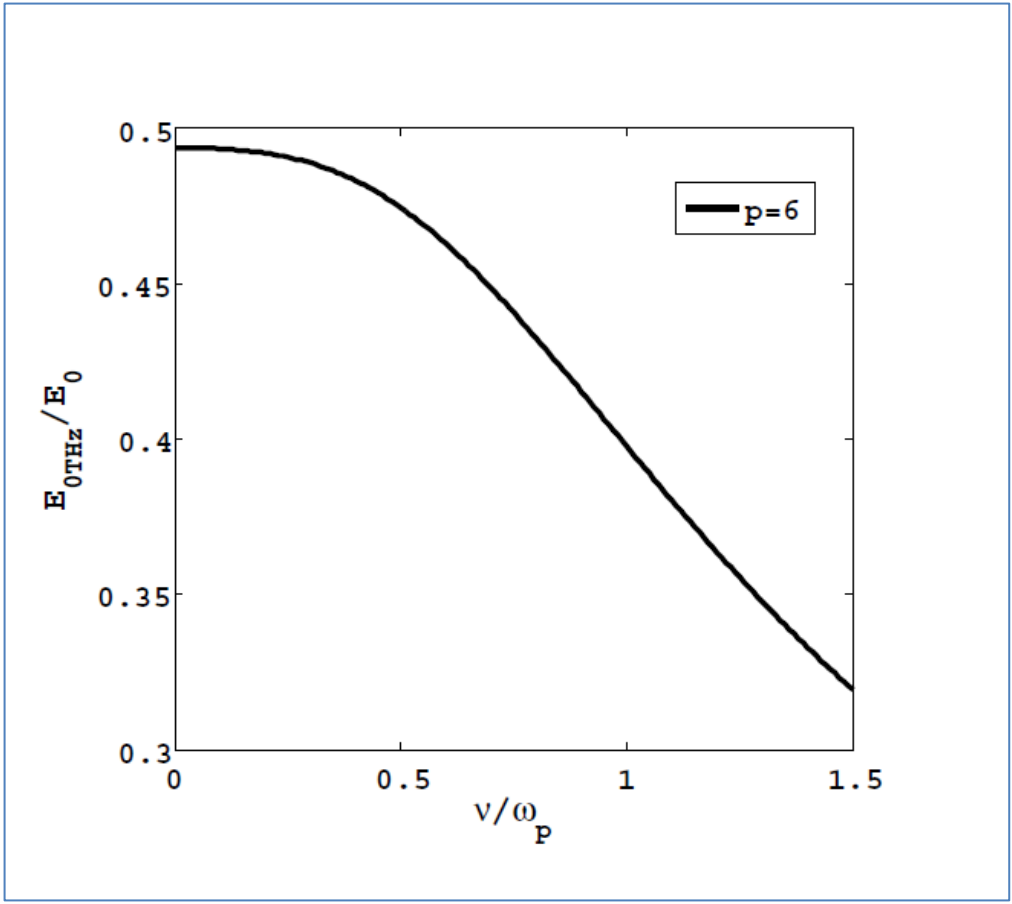

Fig.4. Variation of magnitude of emitted THz radiation field with normalized collision frequency for sG index p = 6 when $b_w = 0.01$ cm, $\omega = 1.15\,\omega_p$, $\omega_p = 2.0\times10^{13}$ rad/sec, $B = 1$ Tesla, $L = 0.5\lambda_p$, $\xi = -0.8L$ and $y = 0.8 b_w$.

In order to clarify the role of magnetic field $B$ and sG index p on the tuning of amplitude and focus of the emitted THz radiation, we have plotted fig. 5. This is clear from the comparison of figs. 5a, 5b and 5c that the effect of magnetic field is to enhance the amplitude of THz radiation. Also the focus of the THz radiation is moved away from the axis of laser propagation. For example, the peaks for the index of p = 6 shifts from about $y/b_w = 0.78$ to $y/b_w = 0.92$ when the magnetic field is increased from 1T to 10T. This observation is the same as realized in another scheme of laser beating [30]. On the other hand, large amplitude and more focused THz radiation is obtained for the case of larger sG index for all the values of magnetic field. Since the ponderomotive force of the laser increases due to the sharp gradient in the laser's intensity for larger index *p*, large current is expected in the plasma that gives rise to the stronger transverse component of the wakefield, which yields the stronger THz radiation.

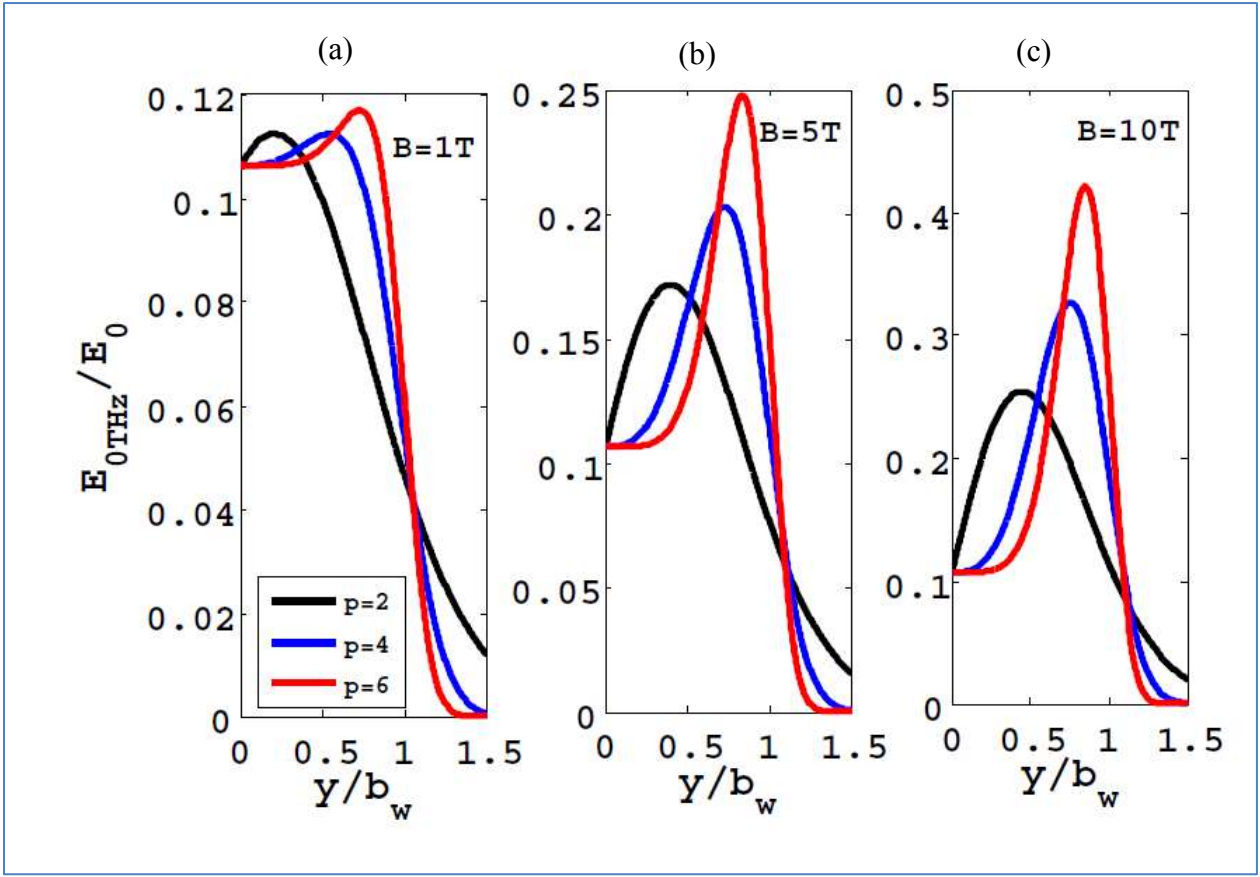

Fig.5. Transverse profile of THz field with normalised distance from beam axis for different magnetic field and sG index, when p = 2, 4, 6, $\omega = 1.45\ \omega_p$, $\nu = 0.5\omega_p$, $b_w = 0.01$ cm, $\omega_p = 2.0\times10^{13}$ rad/sec, $L = 0.5\lambda_p$ and $\xi = -0.8L$.

In order to prove the superiority of sG lasers over Gaussian lasers, we have plotted fig. 6. Here, a comparison of graphs between p = 2 and p = 4 or 6 reveals that higher field of THz radiation is realized for the use of sG lasers. Also, the effect of magnetic field is to enhance this field sharply for the case of p > 2. These results are consistent with the observation made in other schemes with sG and Gaussian lasers [26, 29]. The effect of magnetic field suggests that the stronger THz radiation can be achieved for the case of sG lasers. It means the scheme of THz emission via wakefield could be an efficient scheme if sG lasers are used. However, this is to be mentioned that in both the cases of sG and Gaussian lasers, the maximum amplitude of THz is achieved when the lasers travel a distance of one fourth (or its odd multiples) of the laser pulse length.

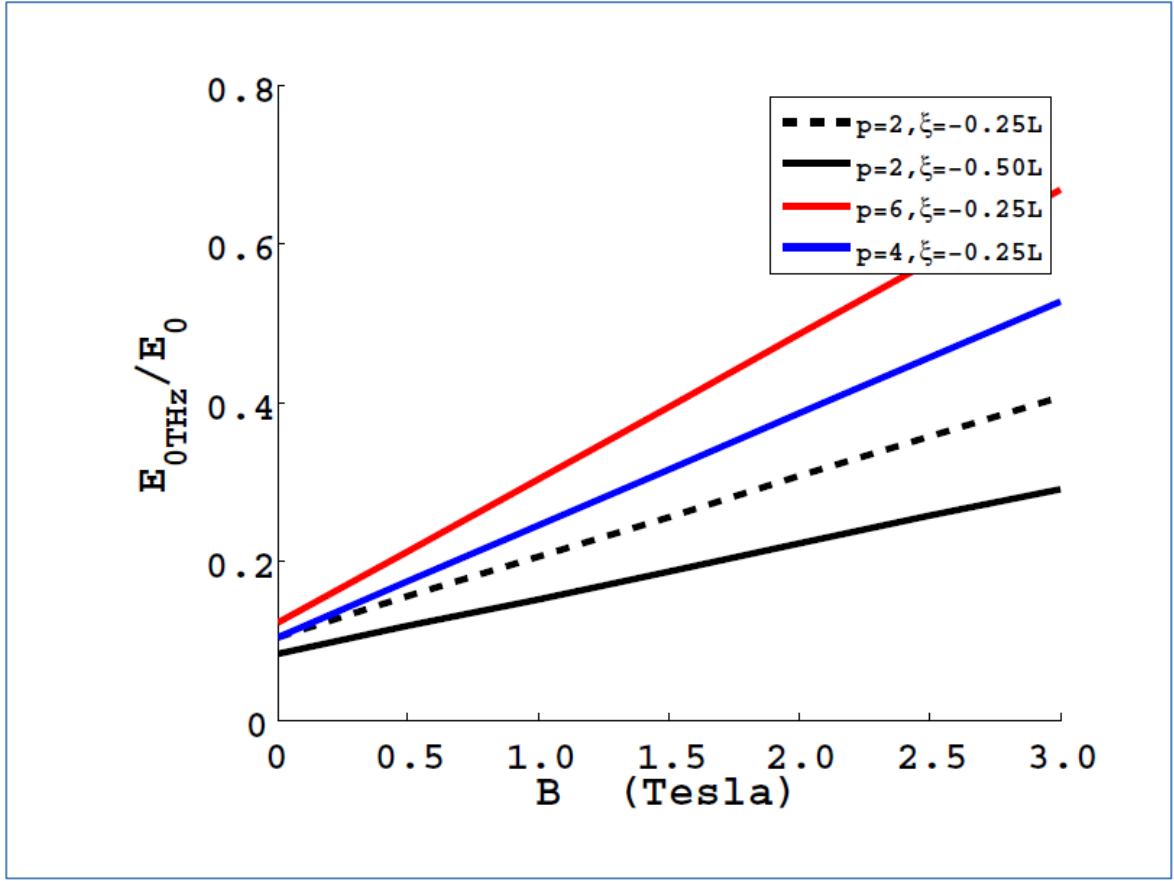

Fig.6. Variation of normalised THz field with transversaly applied external magnetic field for various indices of lasers when $\omega = 2.4\times10^{14}$ rad/sec, $\omega_p = 2.0\times10^{13}$ rad/sec, $v = 0.5\omega_p$ $y = 0.8b_w$ and $L = 0.5\lambda_p$.

Finally, in fig.7, we uncover the contribution of sG index p in enhancing both types of the fields, i.e. the wakefield and the THz radiation. This is observed that the amplitude of THz radiation gets saturated for the case of sG index $p > 6$, whereas the longitudinal component keeps on enhancing under this situation also. However, the rise in THz field with p is faster than that in the wakefield. The most striking result is that the field of THz radiation saturates after

p = 7 and there is no further enhancement in the amplitude. Similar observations were made in another scheme of laser beating [25, 29].

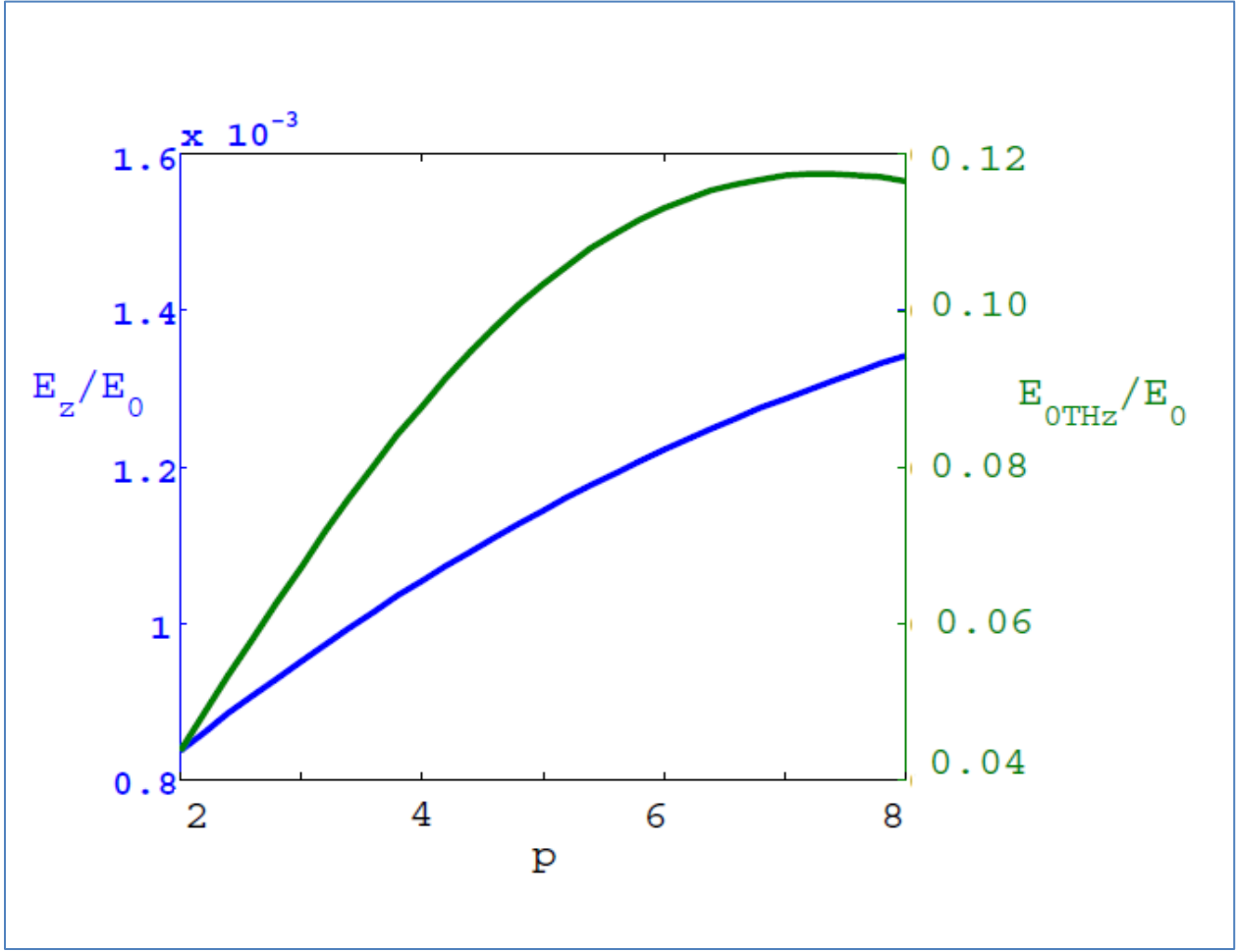

Fig.7. Variation of magnitude of emitted THz radiation and longitudinal wakefield with sG index for $b_w = 0.01$ cm, when $\omega = 1.15\, \omega_p$, $\omega_p = 2.0\times10^{13}$ rad/sec, $B = 1$T, $L = 0.5\lambda_p$, $\nu = 0.05\omega_p$, $\xi = -0.8L$ and $y = 0.8b_w$.

We compare our results obtained with the observations of other researchers. For examples, Hu and his coworkers [29] performed 1D PIC simulations of electromagnetic emission from laser wakefields in magnetized underdense plasmas of density $n_0 = 1.1\times10^{17}$ cm$^{-3}$ corresponding to the plasma frequency $f_p = \omega_p/2\pi = 2.98$ THz, when Gaussian lasers are used. They obtained much lower magnitudes of transverse fields than the longitudinal wakefields.

However, in our scheme fig. 7 clearly depicts that the employment of sG laser of p = 6 enables stronger emission of electromagnetic radiation from laser wakefields. One can also find from fig. 5 that the peak of the transverse field gets sharper than longitudinal field for the same set of parameters. Actually, the nonlinear effects for the case of sG lasers become stronger that make the longitudinal field steep, which results in a sharp increase in its derivative function (see Eq. 13), i.e. the amplitude of the transverse wakefield. Moreover, in our scheme, the effect of magnetic field for the enhancement of THz field amplitude is the similar observation as made by Hu *et al.* [31].

## VI. CONCLUSIONS

The study made on the THz radiation generation in collisional magnetised plasma with the employment of super-Gaussian laser of higher sG index p > 2 clearly depicts that the transverse component of the wakefield gives rise to the electromagnetic fields, which are emitted in the THz frequency range. It is also observed that there is no wake component in the direction of external magnetic field. Collisions between electrons and neutrals not only lead to the decay in amplitude of emitted THz radiation but also limit the focusing of radiation. However, this decay and defocusing may be overcome by the application of an external transverse magnetic field. Employment of lower beam width sG laser over Gaussian laser further enables the tuning of the radiation field amplitude and also enhances the magnitude of emitted THz radiation.

**ACKNOWLEDGMENT**

DST and DRDO, Government of India are gratefully acknowledged for the financial support.